# Arago Optics:
# Maximal Confinement of Traveling Waves

Leizhi Wang[1], Michael Wescott[2], Ming Yin[3], David Tanner[4], Timir Datta[2],

1) Yale University, West Haven, CT 06516

2) University of South Carolina at Columbia, SC 29208

3) Benedict College, Columbia, South Carolina, SC 29204

4) University of Florida at Gainesville, FL 32611

**Motivation & overview**

Optics is limited in the 'ray- approximation' – inclusion of wave properties result in additional phenomena and applications; interferometers and diffraction gratings are two manifestations of such non-geometric, physical optics. Incidentally, the most precise measurement ever, at one part per $10^{21}$ in the (2017) Nobel winning discovery of gravitational waves was achieved with an interferometer. Amendments to the properties of the medium promise negative refractive index meta-materials, perfect imaging, light cloaking, and other ultra-natural marvels. Attention to photon phase, correlations, statistics and wavelength independent phase shifts result in singular optics, quantum optics and anholonomy.

Here we present another possibility, namely 'Arago-optics' to maximize the efficacy of a device by strategically deploying the key qualities along its perimeter. For instance, in conventional sources, waves are generated with maximum intensity at the core; whereas in an Arago-source, intensity is minimal or zero at the center, but highest on villus stretches at the margins. We reason that for a given size and energy output, this radiation profile, produces the highest concentration of energy at the focus, with the maximal confinement of the wave packet. Likewise, the utmost detector resolution is attained when sensitivity is highest on the perimeter and less at the center. This concept holds beyond ultra-focus and Gaussian beams, but generally applies to beams of 'waves' that show constructive and destructive interference. The idea is particularly well suited for a fresh integration of geometry and topology with electronics and materials into real-time wave engineering.

**Introduction & background**

Interest in light must have started way back in prehistory, unfortunately there is little existing record from the Babylonian and Egyptian era. It is from the time of classical Greece that documentation is available,[1] for example in the 5th century before the common era, Plato refers to the phenomenon of refraction; about a century later Euclid applied his deductive and axiomatic genius of geometry to his two books of light – *Optics* and *Catoptrics*. Hero, also of Alexandria in his own *Catoptrics,* inter alia invents the "minimum principle". Without question the classic from the late Early Medieval Period is Ibn al-Haitham or Alhazen's seven-part Optics. Alhazen, like Hero, was a man of practice, he introduced the importance of the plane of reflection and countered Aristotle's idea, of light traveling from the eyes to the object, with the evidence of 'after images' and the proof that sun must be the source of its light, because sunlight can be hurtful to the eyes.

Unsurprisingly, in these earliest manifestations, only rectilinear propagation of light rays was considered and the subject was treated as geometric optics.[2-5] However, some similarities between light and other waves were empirically noted for a long time, and as early as the mid seventeenth century Grimaldi coined the word diffraction. Christiaan Huygens was one of the earliest proponents of the wave nature of light. Remarkably, Isaac Newton was also cognizant of the brightness of the light 'breaking into pieces' near the edges of shadows. Nevertheless, Newton insisted that this 'eel like' behavior is consistent with his own corpuscular theory of light.[6]

The Newtonian view stubbornly dominated optics to the end of the nineteenth century. But the rapid demise of the 'corpuscular theory' followed Young's now famous 'double slit experiment'. Also, the 'Newtonian partisan' Poisson's *reductio ad absurdum* argument against Fresnel's wave theory actually back fired and he Poisson was proven wrong by Dominique Arago's *experimentum crusis* demonstration of the disputed bright spot at the center of a shadow.[7] Now variously known as "Arago-Fresnel-Poisson spot" was previously noted by Grimaldi, but Arago's (re)discovery of it was instrumental for and acceptance of the wave theory of light.[2-5,8]

Optics is limited in the 'ray- approximation' – inclusion of wave properties resulted in additional phenomena and applications; interferometers and diffraction gratings are two manifestations of such non-geometric, physical optics. Incidentally, the most precise measurement ever, at one part per $10^{21}$ in the (2017) Physics Nobel Prize winning discovery of gravitational waves in 2015 was achieved with an interferometer.[9,10]

In the mid nineteenth century with Maxwell's unification of electro-magnetism that visible light was identified with a part of the spectrum of electromagnetic waves. This spectrum ranges from extremely low frequency ELF waves at a few (~ 3) Hz to high energy Gamma rays at ~$3 \times 10^{20}$ Hz, or equivalently wave lengths ranging from several million meters down to ~$1 \times 10^{-12}$ m!

Waves are ubiquitous and amongst the most pleasing spectacles in nature[11-13]; it is no coincidence that the idea of harmony has its origin in waves. As children we have marveled at the beauty of water waves, how they broadcast, move away from the point of origination, reflect off and eventually how the ripples cover up the entire surface. Unsurprisingly this tendency of waves to unfurl led the Renaissance artist and all round genius Leonardo da Vinci to observe, "a wave is never found alone, but is mingled with the other waves". Yes, da Vinci was correct, waves 'mingle with others' - the tendency to spread out is an inherent property of wave propagation.

Waves are also the lowest energy collective excitations and are extended in space and time dimensions (**R**, t); they occur in material media, in vacuum, and even in completely matter-free geometry of space-time, as was recognized by the 2017 Physics Noble Prize.[10] The collective aspect is made visible, when a projectile strikes a body of liquid at rest, viz., still water; in such a case, even though much of the kinetic energy is deposited locally around the point of impact, however the fluid responds to this additional energy by annexing the energy in the form of waves which is distributed over the whole surface. Also, the temporal persistence or the life time of the waves is typically far greater than the duration of the impact, even in a classical system the quality factor Q, can be in excess of a thousand.

A beguiling feature of waves is the simultaneous repetitions in both variables - waves are cyclic in both space and time. Periodicity in space is called the wavelength (λ) whereas the cadence in time is characterized by its frequency ($\nu$) as cycles per second or Hertz (Hz). The space time behavior of a wave is mathematically represented by its wave function W(θ{**R**, t; λ, $\nu$}); here W(θ) is a three parameter function of phase angle θ, the parameters being λ and $\nu$, mentioned earlier and the amplitude $|W|$. Furthermore, θ is modular in both of the space-time variable (**R**, t), namely,

$$W(R, t; \lambda, \nu) \equiv W(\{R \pm i\lambda\}, \{t \pm j\nu^{-1}\}); i \,\&\, j = 0, \pm(1, 2, \ldots integer) \quad \text{1.a}$$

In many situations linearity holds so that the wave field is a solution of the hyperbolic wave equation, namely,

$$\left[\left\{\frac{\partial^2}{\partial x^2} + \frac{\partial^2}{\partial y^2} + \frac{\partial^2}{\partial z^2}\right\} - \frac{\partial^2}{c^2 \partial t^2}\right] W(x, y, z, t; \lambda, \nu) \equiv S(x, y, z, t; \lambda, \nu) \quad \text{1.b}$$

Here, c = λ$\nu$, is the phase velocity. The operator in Eq. 1.b shown (here in Cartesian coordinates) within brackets is the d'Alembertian or wave operator. Equation 1.b is one of the 'holy Trinity' of partial differential equations (PDE); physically, both linearity and PDE are reflections of the field nature of waves. Typically, Eq. 1.b is posed as an initial value problem, that is adequate information about the wave at some initial time is provided and the 'future' wave is to be computed. Interestingly, the number of spatial dimensions in Eq. 1.b influences the nature of the solutions or waves. Consequently, many properties of waves that we are accustomed with are absent in even spatial dimensions.

In our physical 3-dimensional Euclidean space (+1d time) waves follow Huygen's principle, i.e., any where $(x, y, z; t) \in R^3 \times (0, \infty)$ in time and space the function W (solution of 1.b) is uniquely determined by the initial data on the surface of a 'Ball' of radius *ct* and centered at (*x*, *y*, *z*) and the wave equation has progressive solutions that satisfy the following invariance property, i.e., propagate with the fixed rate of c,

$$W(R, t) \equiv W( R \mp ct; \lambda, \nu) \qquad 1.c$$

Notice, c is an explicit in Eq. 1.b and the only parameter that is retained in the solution of the PDE. Also important is that non-linear solitary waves[14] show the above d'Alembertian invariance (1.c), but for the purposes of this article, here we consider only the solutions in the linear limit that obey the principle of waves superposition; hence, the sum of any number of solutions is also a valid solution of Eq. 1.b, i.e.,

$$W(R, t) = \sum_{i=1}^{N} W( R, t)_i \qquad 1.d$$

Where i =1, 2 …N integers, i.e., the resultant W(**R**, t) or the outcome of N waves; mathematically, sum of the solutions of Eq.1.b is also a solution. Also, since here 'adding displacement' addition of waves is similar to vector addition.

This equation of linear superposition principle is central in interference and diffraction of waves and will be used in this article. Physically equation 1.d implies that even though many waves can be present, but at any given space-time (**R**, t) point a single, or one and only wave W(**R**, t) may exist.

**Arago optics**

Here we use the term Arago optics in a broader sense than in the past: first (1) include a wide range of waves, secondly (2) situations in addition to, and not just blocking (shadowing) of waves and finally (3) to distinguish from conventional Fresnel optics. Our arguments apply to any cyclic process where the principal of superposition results in stationary patterns of constructive and destructive interference.

The tendency of spreading out (mentioned earlier) of waves in open space, arise from the time evolution of waves (Eq.1.b) and is mathematically stated in several ways if at initial time the wave is nonzero over a small region of space, then W will become non-zero over all space and time. Physically as the wave travels forward, this process distributes the total energy over the ever increasing volume, thereby decreasing the density and hence the locally available energy. In presence of obstructions the wave can produce a wide variety of scattering phenomena.[15] Also, in text books such as in reference 2 and others the interplay between one wave and another, or with the wavelets of its own are distinctly classified as either interference or diffraction, however in reality, many natural phenomena violate such categorization.[16,17]

It is of great significance to determine the design for a given system, that focuses the highest concentration of wave energy. Our strategy relies on equation 1; hence, it is important to maintaining linearity; non-linear effects including tsunami, solitary waves, sonic-bullets,[18] KdV or Sine-Gordon waves are excluded.[14] The present challenge is to deliver a maximally confined wave packets using traveling wave trains.

We analyze the features critical in designing Arago optics. In particular, the emphasis of this article will be on sources that would give rise to the most spatially compact 'spot' or focus. Traditionally sources in Gaussian optics, as a matter of fact often even non-Gaussian beams [19-21] are compact, waves start at an intense origin and subsequently with appropriate 'optics', care is taken to create the desired beam downstream. In the past, a light source that is brightest at middle was natural; because light used to be produced by chemical combustion or thermal radiation which of necessity requires a core. However, with the advent of light emitting diodes and lasers it is feasible to design sources with non-traditional geometry, in the visible or even shorter wavelengths.

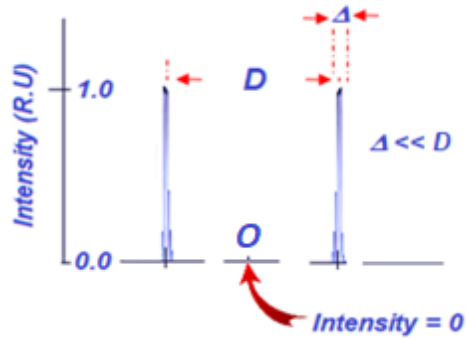

**Figure 1.** The intensity profile of an ideal Arago source is shown. Concentrated radiation from the margin and the characteristic zero output at the origin O is high-lighted.

In this article our concern will be on sources of waves. The sources of interest are those with strongest intensity output at the periphery and no emission from the origin. Emphasis on zero output at the center O but high output from slender edge regions distinguishes the proposed idea from currently available wave sources and is the hallmark of our design. The ideal intensity distribution of an Arago source is shown in Fig. 1.

Notice that in figure 1 the radiation distribution is not monotonic, the intensity is highest and localized ($\Delta$) in the peripheral regions. The figure also indicates the signature mark of the concept, namely minimal intensity over a region of diameter D (D>> $\Delta$) around the origin or center of the source. The geometric argument in favor of large D is shown in figure 2, below. Consider a toy model with coherent waves originating from point sources located at A, B and C, D, (on the object plane) with separations d′ and d respectively.

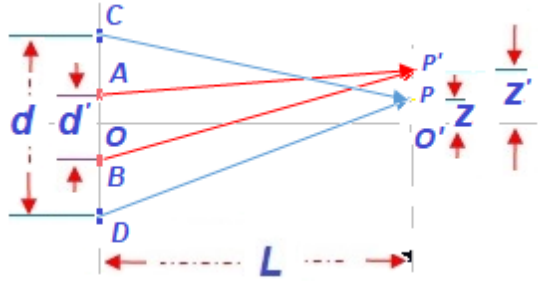

**Figure 2.** The positions of the 1st minimum P (P′) for two pairs of waves, following paths AP′-BP′ and CP-DP are shown (not to scale). The point O′ lies on the locus of the perpendicular bisector of the line CD. With the separation between the sources AB and CD as d′ and d respectively and d>d′, the HW of the central intensity peak follows z<z′.

At point O′, the center of the image plane, by symmetry both pairs of waves mutually superpose constructively and give rise to a maximum. As a matter of fact, O′ is the position of the zeroth order peak; it is a maximum for all values of the separation d and is independent of the wavelength. However, because the paths differences Δl for all other points on the image plane are different, the locations of other order interferences, most importantly, the position of the first minimum that defines half width (HW) of the central intensity peak are not the same. The waves from A and B produce destructive interference at P′ and those from C and D at P. In the far field approximation applying the condition of destructive interference at O′P it can be shown that, the half width or distance z on the image plane, depends inversely on the source separation (d), and one obtains

$$\frac{z}{\lambda} \sim \frac{L}{d} \qquad \text{2.a}$$

Also, since d>d′ it follows

$$z \sim \lambda \frac{L}{d} < z' \sim \lambda \frac{L}{d'} \qquad \text{2.b}$$

Equation 2.b, confirms the earlier stipulation that a tighter spot with a narrower focus results when the interfering waves emanate with larger initial separation.

For a spot with azimuthal symmetry along the propagation axis, emitter is required to possess cylindrical symmetry about the said axis. In such situations the radiating surface will be of multiply connected topology and the source be in the shape of a ring. Furthermore, for two ring shaped sources, one twice the radius of the other, radiating waves with the same power, the larger source will focus energy over a sharper central maximum ideally with twice peak height and half the peak width compared with the smaller source.[8]

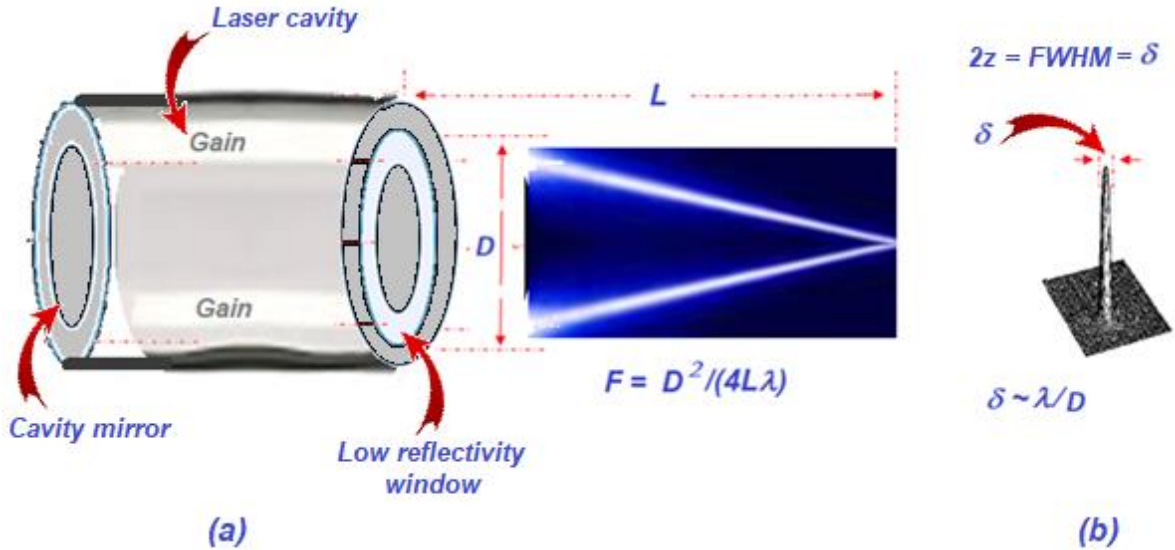

**Figure 3:** A (not to scale) semi-cut out view of an Arago emitter with a gain medium. (a) Cylindrical-shell cavity with a ring shaped exit port along with the outgoing waves are shown. (b) Sketch of the spot at the focus.

A bright ring-shaped source can be constructed from a laser, such as one with a cylindrical-shell shaped laser cavity is shown in figure 3. Notice, that in this configuration, the central requirement of zero intensity over the center region of the source is conveniently achieved by having low reflectivity only over the circular region of size D on the output side of the cavity.

A source with multiple sets of emitters is also possible. If the radiator comprises of multiple (rings of) emitting zones, then stationary interference can be guaranteed when the distances along successive optical paths increase consecutively by one wavelength or phase by $2\pi$. N, the number of radiators determines the maximum intensity $I_{max}$ of the central peak, and for perfect constructive interference, $I_{max} \sim N^2$ and Half width $\sim N^{-1}$. [2-5, 8]

For instance, the case of a source with N circular rings is shown in figure 3. Let us frame the interference in terms of $\Delta\phi j$, the phase of the wavelets arriving at the focus from the source zone 'j'. If the phases along successive optical paths 1O′, 2O′, … NO′ consecutively increase by $2\pi$ then all the wavelets constructively superpose and give rise to a central peak at the focus at O′. Under this condition, the resultant wave in the region of the diffractive peak will be maximally confined in space with the corresponding full width of half maximum (FWHM) or peak $\delta$, given by,

$$\delta \approx \frac{\lambda L}{4a} = \left(\frac{\lambda}{4D}\right) L \qquad \qquad 3.a$$

Notice, that equation 3.a follows from the assumption of perfect transvers symmetry plus that all wavelets are exactly identical. Hence, does not contain any errors or influence of imperfections; however, in practice the quality of the focus is critically dependent on the irregularities in the magnitudes and phases of the interfering wavelets. In practice uniformity of geometry and fluctuations in the gain medium are limiting factors.

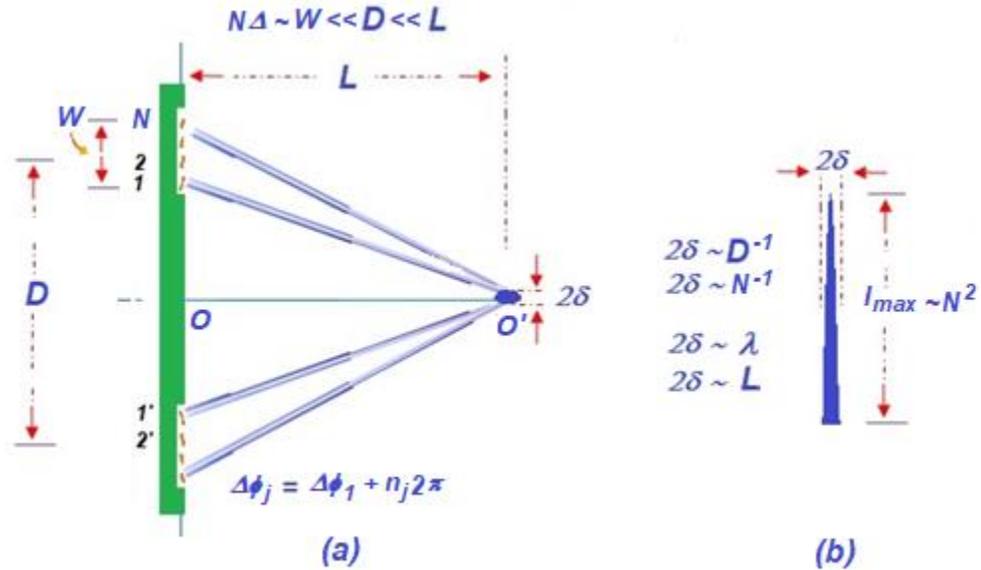

**Figure 4:** A (not to scale) diagram of an Arago source with N sets of peripheral emitters and no emittance over the central disk of size D is shown. (a) All wavelets combine constructively when successive phases differ by integer ($n_j$) multiples of $2\pi$. For clarity only two pairs of paths are drawn, point O' is the 1$^{st}$ order focus at a distance L from the center O. (b) Shows the dependence of the intensity and full width of the central peak, on the parameters D, N, L and $\lambda$.

To estimate of the effect of irregularities in the geometry of the emitting zones a critical figure of merit is the ratio $\varepsilon/\Delta$, which determined by the tolerance of the fabrication lithography. Consequently, the quality of confinement depends on satisfying the following condition,

$$\frac{\epsilon}{\Delta} < \frac{\Delta}{\lambda} < \frac{L}{D} \qquad \text{3.b}$$

If the tolerance is large so that the of the edges of emitters are uneven with significant random dispersion of $\Delta$, then the spot geometry quality will be degraded and not be satisfactory. In the visible spectrum this would imply fabrication precision with tolerance well into the submicron regime. Also, these villus source regions need not be continuous nor change the topology of the emitting surface; however, abrupt termination points may give rise to additional (unwanted) diffraction. In practice the diffraction spot is extremely sensitive to axial misalignment and imperfections, irregularities, at the source and the edges. Two examples of which are shown in figure 5.

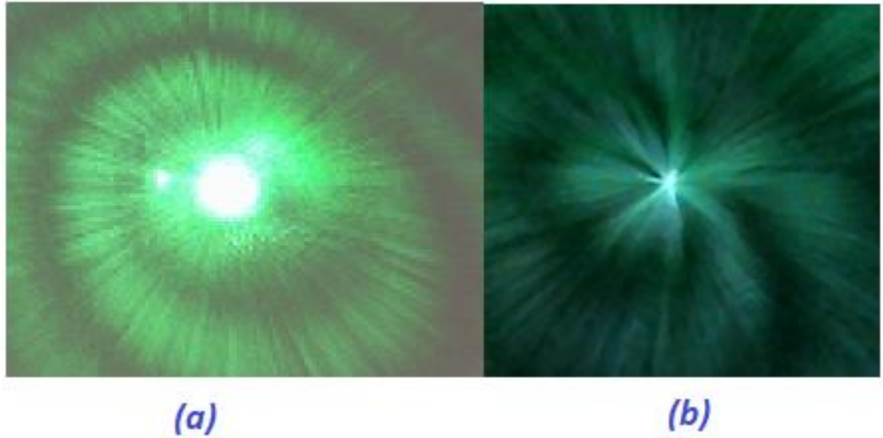

**Figure 5:** Sensitivity of the central (Arago) diffraction peak to real world complications. (a) A well-formed zeroth order spot with slight misalignment. (b) Effects of laser imperfections and azimuthal index.

Often it is useful to define the Fresnel number of the problem, F is determined as follows,

$$F = D^2 \left(\frac{1}{4L\lambda}\right) \qquad 4.a$$

F is needed in estimating which Fresnel or Fraunhofer limiting field conditions hold. [2-5] In terms of F, $\delta$ the width of the central diffraction peak is given by

$$\delta = \left(\frac{D}{F}\right) \qquad 4.b$$

In Eq. 4.b the inverse D dependence of $\delta$ is obscured, because D is explicit as the numerator. This gives the superficial impression of linear dependence of d with D, however in reality as shown in Eq. 4.a, the denominator Fresnel number is quadratic in D so the correct reciprocal relationship with $\delta$ is ensured.

The beam intensity can be further optimized, by suppressing destructive interference amongst the emergent waves. This is achieved with a set of alternate Fresnel zones at the exit port of the Arago laser. For audible and low frequency waves these Fresnel structures can be electro-mechanical 'speakers' at ultrasonic frequencies they can be piezoelectric generators. Assuming the intensity from a single zone to be given by, I ~$I_0$, then a set of N zones with the same dimension will ideally give rise to a radiation intensity I ~ $N^2$. In all cases the emitting regions can be fabricated as a set of concentric hoops, arranged in the form of a spherical cap. Further benefits will follow by reduced shadowing, blazing and controlling the output from successive hoops.

In figure 3, an arrangement utilizing a gain region was sketched. The gain medium can be any of the established and proven mixtures of gasses; crystalline solid, or an amorphous or glassy medium; as a matter of fact, many high-power lasers use Neodymium glass.[22] For gain in the optical regime the concerns are to stay away from absorption and promote non-linear effects that amplifies the radiation, also the proximity of metallic conductors, semiconductors and junctions are all factors that influence the electromagnetic field, hence may provide additional handle on the radiated beam.[23-28] For Self Amplified Spontaneous Emission (SASE) and collective non-linear instability in Free-Electro-Lasers (FEL) the cylindrical shell region (fig. 3) may comprise of the undulator magnets in a wiggler of an X-ray laser. [29-32]

The topic of wave confinement is of topical interest, [33-34] here we have discussed the generation of maximally focused waves, but similar considerations apply for imaging elements as well. From the interference point of view both Arago and Fresnel lenses are similar, but in Arago the emphasis on the extremal regions. Also, with the advance in fast electronics and new materials, it may soon be feasible to alter and control in real time, the parameters of the Arago elements, hence they can be dynamic and not plastic. It may be noted that other proposed applications of Arago's spots such as NASA's 'aragoscope'[35] are not dynamic and utilizes one single central obstruction hence lack the multiple zones superposition described here. In addition, the 'optically active part' of a single obstruction (as in the proposed design) is merely the ~$\lambda$ wide peripheral region, hence the total flux of waves received will be low. Instead, we reason that a far higher 'return on the NASA investment' will result, if a cascade of dioptric and or catoptric elements are introduced around the circumference of the obstructing disk.

In a previous section we chose to discuss the interference condition directly by the phases $\Delta\phi j$ of the waves, instead of the more common optical path lengths. Because, consideration of the phases $\Delta\phi j$, allow the possibility of active dynamical controllability. With current state of technology, at the relevant time scale or frequency for many operational values of the distances $\lambda$, W, D and L, it may be possible to massage the initial phase of the radiated wavelet at the source. By such manipulation the location of point O as well as the width of the peak may be regulated as desired.

The importance of "wave-making drag or resistance" is not generally appreciated, it is surprising that in a mechanical medium there is a cost to create waves[13]; but, even source-free, plane electromagnetic waves propagating in the emptiness of perfect vacuum space are subjected to $Z_0$ = 119.917 ohms of impedance of free-space. More surprisingly, also in energy conserving Hamiltonian systems of quantum mechanics, there is an impedance to the Schrodinger waves.[36] Consequently, in designing the optimal source for a particular type of wave, impedance matching will be technologically important. Furthermore, many observable effect are due to fluctuations in the field, plus are extremely sensitive to the geometry and physical properties of the boundaries, material parameters such as conductivity, band gap, nature of the edge states may all be key factors in the optics of the future.

In many ways, at present wave optics has over taken geometric optics. This is partly due to the fact that new phenomena beyond classical amplitude superposition and positive refraction, have also been discovered or proposed. As alluded to previously, statistical effects like Hanbury Brown-Twiss type photon bunching or anti-bunching plus the possibilities of negative refractive index systems and others have opened up new opportunities in many fields including quantum optics, photonics and meta-materials.[37-42] The concept described here is likely to extend beams beyond Gaussian waves, and investigate novel concepts such as, holonomy, vortex-waves, wave-bullets, singular and diffraction-free optics.[43,44]

Great progress has been made in adaptive optics.[45-48] However, the configuration of most optical devices is plastic, that is not readily changeable. The idea presented above is particularly well suited for a fresh approach to 'elastic' optics. Because each device comprises of functionality only along its outer most boundary, the number of addressable units scale linearly with dimension (L), in contrast a mirror or lens scales at least $\sim L^2$. Hence real time electronic manipulations of an Arago optical device will be far nimbler. We argue that in integrating fast adoptive electronics to new materials with controllable geometry and topology, the present idea provides a clearer path leading to dynamically changeable, elastic or 'chameleon optics' with far superior control over the waves.

**Potential Applications**

We anticipate that maximally compact waves, especially in conjunction with optical elements that are 'elastic' (in the sense defined above) will be important in a wide range of technologies including health care and life sciences such as non-surgical medical procedures such as "shock wave lithotripsy", lipolysis, skin and surface treatments; in far-field and super-resolution microscopy. In the physic, engineering and medical applications high-intensity radiation spots are in demand in applications such as removal of debris in outer space and ignition initiating in inertially confined fusion. A short list of likely areas of in science and technology, is as follows,

- Acoustics
- Airy/Bessel beams
- Beam focusing
- Imaging
- Interferometry
- Lighting
- Ultra-resolution lithography
- Manufacturing
- Medical procedures
- Microscopy
- Particle control/manipulation

- Spot/focus control
- Sound control/manipulation
- Inertial confinement of Fusion
- Ultrasonics
- Intense X-rays

## Summary

The most notable recent discovery of gravitational waves is attributable to advances in wave optics. Here we have considered parameters necessary - within the linear regime- to focus waves of a given wavelength $\lambda$. into the most compact packets, that is to attain the best or maximally confined wave with the smallest width $\delta$, of the intensity distribution. We utilize the principle of (wave) superposition. The essential distinction of the proposed Arago optics from the conventional (including the Fresnel) types, is the emphasis on the edges, intensity of a source is highest at the boundary but lowest or zero at the center. Our concept is applicable to all waves, including beams of non-Gaussian - viz., Bessel, Airy and others- and hence is particularly useful for situations where lenses are not practical, for example, ultra-sound, microwave, free-electron lasers and X-rays. Because the proposed device comprises of functionality only along its outer most boundary the number of addressable units scale linearly with dimension the concept is singularly nimble for real time manipulations. Incorporating unconventional configurations of new materials as controllable elements is an attractive path to elastic or 'chameleon optics' of the future.

## Acknowledgments

This work was partially supported by the University of South Carolina and the University of Florida. One of us (TD) would like to dedicate this article to the inspiring interactions and fond memories of Maximilian Jacob (Max) Herzberger.[47]

## References

1) Max Herzberger, Applied Optics, **5**, 1383 (1966)
2) Eugene Hecht, Optics, 5$^{th}$ ed., ISBN-10:0133977226, Addison-Wesley (2002)
3) R. W.Ditchburn, *Light*, ISBN-10:0486666670, Dover (1991)
4) Max Born and Emil Wolf, *Principles of Optics*, 7$^{th}$ ed, ISBN-10:0521642221 Pregamon (2002)
5) Arnold Sommerfeld, *Optics*, Lectures on Theoretical Physics vol 4, Academic (1954)
6) Isaac Newton, *Optics*, London S. Smith and B Walford (1704)
7) Pascal Fischer et al, Optics Express, **15**, 11872 (2007)

8) Timir Datta, Physica Scripta, **90**, 038002 (2015)
9) Chris L Mueller et al., Review of Scientific Instruments **87**, 014502 DOI:10.1063/1.4936974 (2016)
10) https://www.nobelprize.org/nobel_prizes/physics/laureates/2017/press.html
11) A. P. French, *Vibrations and Waves*, Norton (1971)
12) Stephen W Hawking, *A Brief History of Time*, ISBN:0-553-05340-X, Bantam (1988)
13) Horace Lamb, *Hydrodynamics*, 6$^{th}$ ed, ISBN:0-486-60256-7, Dover (1945)
14) Willy Hereman, arXiv:1308.5383v1 (2013)
15) Timir Datta, et al., arXiv:0901.0140v1 (2008)
16) R.P. Feynman & R. B Leighton & M. Sands, *The Feynman Lectures on Physics*, Vol 1, Addison-Wesley (2007)
17) T. Datta J Thermodynam Cat 3: e112. doi:10.4172/2157-7544.1000e112 (2016)
18) Alessandro Spadoni and Chiara Daraio, PNAS, **107** 7230, DOI:10.1073/pnas.1001514107 (2010)
19) J. Durnin, et.al, Phys. Rev. Lett. **58**, 1499 (1987)
20) M. V. Berry J. Opt. A: Pure Appl. Opt. **6**, 259 (2004)
21) Julio C Gutiérrez-Vega and Carlos López-Mariscal, J. Opt. A: Pure Appl. Opt.10-015009 (2008)
22) J. E. Geusic, H. M. Marcos, *and* L. G. Van Uitert, Appl. Phys. Lett. **4**, 182 (1964); doi.org/10.1063/1.1753928
23) R. S. Craxton et al., Physics of Plasmas, **22**, 110501 (2015); doi: 10.1063/1.4934714 (2015)
24) K. I. Golden, et al., Physical Review A **11**, 2147 (1975)
25) T. Datta, M. Silver, Applied Physics Letters **38**, 903 (1981)
26) T. Datta, J. A. Woollam, Physical Review B **39**, 1953 (1989)
27) Anika Kinkhabwala, et al, Nature photonics, DOI: 10.1038/NPHOTON2009.187 (2009)
28) T. Datta, L. H. Ford, Physics Letters A **83**, 314-316 (1981)
29) Miguel A. Bandres et al., Science, **359**, 1231 (2018)
30) G.A. Siviloglou et al., DOI: 10.1103/PhysRevLett.99.213901 (2007)
31) D.L. Matthews et al., Physical Rev. Letts., **54**, 110 (1985)
32) C. Pellegrini, Physica Scripta, 014004 (2016)
33) Sanematsu, Paula Cysneiros, "Propagation of Periodic Waves Using Wave Confinement. " Master's Thesis, University of Tennessee (201) http://trace.tennessee.edu/utk_gradthes/747
34) John Steinhoff, Andrew Wilson, Subhashinni Chitta, and Frank Caradonna. "Wave Confinement: Long-Distance Acoustics Propagation through Realistic Environments", 20th AIAA/CEAS Aeroacoustics Conference, AIAA AVIATION Forum, DOI:10.2514/6.2014-3196
35) www.nasa.gov/content/the-aragoscope-ultra-high-resolution-optics-at-low-cost
36) T. Datta, R Tsu, arXiv preprint cond-mat/0311479 (2003)
37) Gordon Baym, arXiv:9804026v2 (1998)
38) Ian Walmsley, Science, **358**,1001, DOI: 10.1126/science. aao3883 (2017)

39) J.B. Pendry, Phys. Rev. Letts., **85**, 3966 (2000)
40) Jordi Bonache et al., Scientific Reports, DOI: 10.1038/srep37739 (2016)
41) Weiguo Yang, Michael A. Fidy, arXiv:1306.2351v1 (2013)
42) J.B. Pendry et al., Science **358**, 915–917 (2017)
43) M.V. Berry, Nature, **326**, 277 (1987)
44) M V Berry J. Opt. A: Pure Appl. Opt. **6**, 259 (2004)
45) Na Ji, et.al., Nature Methods, **7**, 141 DOI:10.1038/nmeth.1411(2010)
46) Alastair Basden et al., arXiv:1010.3209v1 (2010)
47) http://www-groups.dcs.st-and.ac.uk/history/Biographies/Herzberger.html
48) Max Herzberger, Modern geometrical optics, R.E. Krieger Pub. (1980)